\documentclass{iau}

\usepackage{amsmath}
\usepackage{graphicx}
\usepackage{multirow}
\usepackage{subfigure}

\begin{document}

\lefttitle{Matteo Cerruti}
\righttitle{Unsupervised Classification of GRBBLs with Machine Learning}

\jnlPage{1}{4}
\jnlDoiYr{2025}
\doival{10.1017/xxxxx}

\aopheadtitle{Proceedings IAU Symposium UniversAI}
\editors{Editors: xxx}

\title{Unsupervised Classification of Gamma-ray Bursts from Blazars (GRBBLs) with Machine Learning }

\author{Matteo Cerruti}
\affiliation{Université Paris Cité, CNRS, Astroparticule et Cosmologie, F-75013 Paris, France \\ 
              \email{cerruti@apc.in2p3.fr}}

\begin{abstract}
Blazars dominate the extragalactic $\gamma$-ray sky and show pronounced flares. Using public Fermi-LAT light curves for 732 blazars with secure redshifts, I implement an automated pipeline to identify and characterize $\gamma$-ray bursts from blazars (GRBBLs). Each event is modeled with an exponential rise/decay profile, and spectral variability is quantified via a constant fit. From 679 high-quality GRBBLs, I apply extreme deconvolution for unsupervised classification. The GRBBL population is remarkably homogeneous; the most robust split is in achromatic vs. chromatic events, with significant overlap. Removing spectral information yields a luminosity-driven classification in type-1 and type-2 GRBBLs, although this classification is not identified in all tests. This study establishes GRBBL population studies as a tool to study blazars. As a by-product of this project I identify a correlation between peak luminosity and timescales in GRBBLs.
\end{abstract}


\maketitle

\section{Introduction}

Blazars are the most common extragalactic objects in $\gamma$-rays. Their emission spans the full electromagnetic spectrum, from radio to very-high-energy $\gamma$-rays, and is characterized by a high degree of polarization (in radio, optical, and X-rays), and extreme variability, sometimes down to timescales of minutes. Within the unified model of active galactic nuclei (AGNs), they are understood as radio-loud AGNs whose relativistic jets are aligned with our line of sight, boosting their non-thermal emission.
The Fermi-LAT instrument \citep{fermilat} has revolutionized our view of blazars by providing uninterrupted light curves for thousands of sources over more than 15 years of operations. This wealth of data has triggered numerous studies of blazar variability, often in the context of multi-wavelength and, more recently, multi-messenger campaigns. However, these analyses almost always target individual objects or small samples selected for specific properties. There has been, until now, no systematic population study of blazar flares: no attempt to characterize their global properties, investigate the parameter space they occupy, or explore possible classes within the population.
I propose to refer to these events as gamma-ray bursts from blazars, or GRBBLs, a simple acronym that avoids the awkward periphrases that have been used in the literature. The goals of this project are threefold: (i) to provide an automated identification of all GRBBLs detected by LAT in sources with a known redshift; (ii) to characterize their observational properties; (iii) to explore whether the population exhibits natural groupings or classes, using unsupervised classification techniques.
This conference contribution represents a summary of the results published in \citet{grbbl}.

\section{Automated Identification and Characterization of GRBBLs}

\begin{figure}[t!]
        \centering
        \begin{subfigure}{}
            \includegraphics[width=0.45\textwidth]{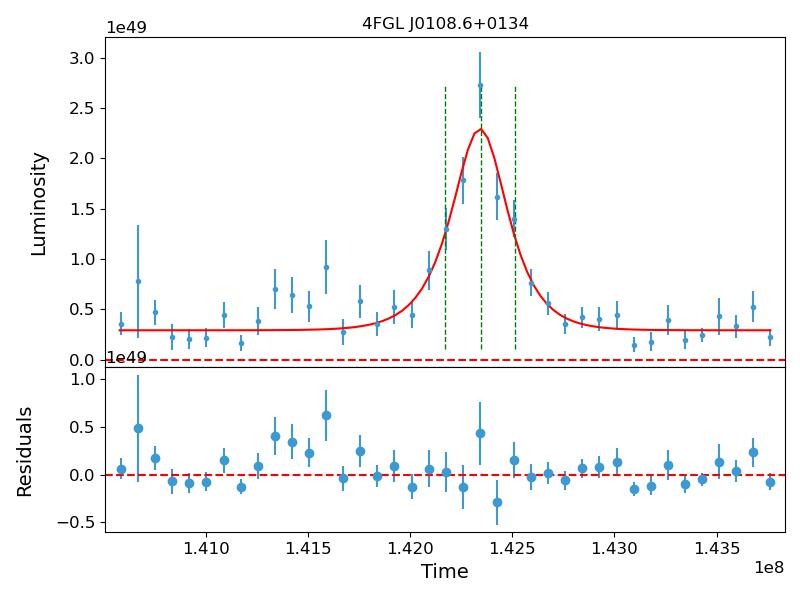}
        \end{subfigure}
        \hfill
        \begin{subfigure}{}
            \includegraphics[width=0.45\textwidth]{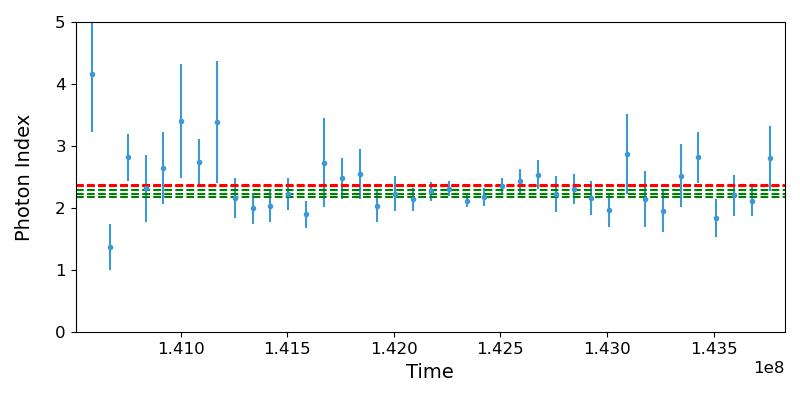}
        \end{subfigure}
        \caption{A GRBBL identified in 4FGL~J0108.6+0134, from \citet{grbbl}. Left top: Luminosity vs time (the vertical green lines indicate $t_0$ (the central one), $t_0 - 2\tau_r$ (on the left), and $t_0 + 2\tau_d$ (on the right); the red line represents the fit function); Left bottom: Residuals vs. time; Right: Photon index vs. time (the horizontal green lines show the best-fit value of $\Gamma$ plus or minus its uncertainty; the horizontal red lines show the same but for $\Gamma_{4LAC}$).}
        \label{figone}
    \end{figure}

The analysis starts from the public LAT light-curve repository \citep{LCR}. From the 4LAC-DR3 catalog, I selected all sources classified as blazars, with available redshift measurements. After a careful manual vetting of the redshift references (see the CDS tables at \url{https://cdsarc.cds.unistra.fr/viz-bin/cat/J/A+A/698/A101} for full details on the source selection and redshift cleaning), the final list contains 732 blazars. For each source, I retrieved light curves binned at 3-day, 7-day, and 30-day intervals, including both the integrated energy flux and photon index. The time axis was corrected for redshift (dividing by 1+z) and fluxes were converted to luminosities using standard $\Lambda$CDM cosmology.
Light curves often contain spurious points due to non-converging likelihood fits. These outliers were removed through quality cuts on flux, photon index, and flux-to-TS ratio, ensuring clean, high-quality time series.
The identification of GRBBLs was done through a fully automated two-step procedure. In the first step, I applied the Bayesian blocks algorithm \citep{scargle13} to segment the light curve, using a false-alarm probability $p_{BB} = 0.1$ for the fiducial dataset, with tests at $p_{BB} = 0.01$ and $p_{BB} = 0.001$ for robustness. In the second step, I merged consecutive blocks into superblocks, ensuring that each segment contains a unique luminosity maximum. By construction, each superblock defines a candidate GRBBL interval.\\

For each event, I fitted a simple exponential rise–decay profile,

\begin{equation}
L(t) = L_{base} + \frac{L_{peak}}{2^{-\frac{t-t_0}{\tau_r}} + 2^{+\frac{t-t_0}{\tau_d}}}
\end{equation}

with five free parameters: the two timescales $\tau_r$ and $\tau_d$; $t_0$ and $L_{peak}$, which are respectively, the time and luminosity where the two exponential functions meet; and  $L_{base}$, which is simply a constant baseline flux.
Spectral variability during the event was analyzed in parallel. For each GRBBL, I extracted the photon index evolution within a window ($t_{eff}$) from $t_0 - X\tau_r$ to $t_0 + X\tau_d$ (where X = 2 in the fiducial test), and fitted it with a constant model. This provides the average photon index $\Gamma$ and the reduced chi-2 value $\tilde{\chi^2_\Gamma}$ that quantifies spectral evolution: low values indicate achromatic GRBBLs, while high values flag chromatic ones. An example of a GRBBL, with its fitted function, is shown in Fig. \ref{figone}.\\

Strict quality filters were then applied. Light-curves with missing bins within $t_{eff}$ were rejected; only those with convergent fits, well-defined timescales (at X$\sigma$, X = 2 in the fiducial test), and individual residuals within 3$\sigma$ were kept. Across the three binnings, duplicate identifications were resolved by keeping the fit with the smallest relative parameter uncertainties.
The resulting fiducial sample contains 679 high-quality GRBBLs. For each, I stored eight parameters: $L_{peak}$, $\tau_r$, and $\tau_d$ from the light curve fit, together with an explicit asymmetry parameter, $a = (\tau_d - \tau_r)/(\tau_d + \tau_r)$ ; $\Gamma$ and $\tilde{\chi^2_\Gamma}$ from the photon index fit; and lastly the average values of luminosity and photon index from the 4LAC catalog, $L_{4LAC}$, and $\Gamma_{4LAC}$.

\section{Unsupervised Classification}

\begin{figure}[t!]
        \begin{subfigure}{}
            \includegraphics[width=0.45\textwidth]{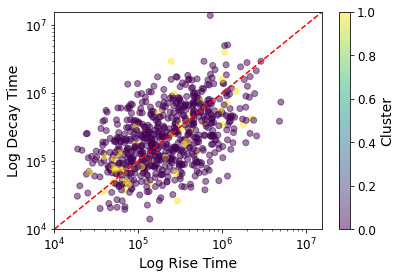}
        \end{subfigure}
               \hfill
        \begin{subfigure}{}
            \includegraphics[width=0.45\textwidth]{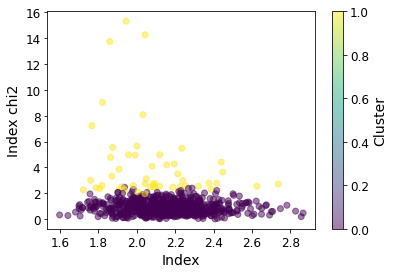}
        \end{subfigure}
        \caption{Results for the Extreme Deconvolution algorithm run on eight parameters. Top: correlation plot for $\tau_d$ vs $\tau_r$ (error bars have been removed for clarity) color-coded by the most likely cluster for $k=2$; same as before, but for the $\Gamma$ vs $\tilde{\chi^2_\Gamma}$ correlation. Figure taken from \citet{grbbl}.}
        \label{figtwo}
    \end{figure}

To explore whether natural clusters exist within the GRBBL population, I applied unsupervised classification methods in the eight-dimensional parameter space. As a first test, I used Gaussian Mixtures (scikit-learn), after log-transforming luminosities and timescales and normalizing all variables. This approach, while illustrative, does not account for uncertainties and tends to overfit: the BIC (Bayesian information criterion) suggested that the optimal number of clusters is k=6, but the resulting clusters mainly reflected asymmetry, with no clear physical meaning.
I then applied Extreme Deconvolution \citep{Bovy}, which incorporates measurement uncertainties. With the fiducial parameter set ($p_{BB} = 0.1$ and $X = 2$), the optimal classification yields two overlapping clusters, driven by the photon-index evolution (Fig. 2): achromatic GRBBLs (the majority), with a photon index consistent with a constant during $t_{eff}$, and chromatic GRBBLs, showing significant spectral changes during $t_{eff}$ (see Fig. \ref{figtwo}). Removing the spectral fit quality information ($\tilde{\chi^2_\Gamma}$), and working with the 7 remaining free parameters, results in a luminosity-driven classification, with type-1 GRBBLs with high-L$_{peak}$ and small dispersion, and a second population of type-2 GRBBLs with a much larger dispersion, explaining the low luminosity events.\\
This result depends of course on the hyper-parameters of the algorithm and in particular on $p_{BB}$, used to automatically segment the light-curves, and on $X$ that controls both the duration of the GRBBL and the quality of the fit. The results from the different combinations of  $p_{BB}$ and $X$ show that the classification of GRBBLs in chromatic and achromatic is solid, while the classification in type-1 and type-2 disappears  for larger values of X and smaller values of $p_{BB}$. For the full details on these additional tests, see \citet{grbbl}. \\

\section{Discussion and Conclusions}
The key result is the remarkable homogeneity of the GRBBL population. There are no well separated clusters, nor outliers. The clearest classification is in chromatic/achromatic GRBBLs, but the transition is smooth and with significant overlap. Removing the information on the spectral variability, there is an evidence for a classification in type-1 and type-2 events, although this classification is not seen in all tests. The homogeneity of the sample is an important piece of information for theoretical models, that should be able to reproduce all events with a smooth transition in their parameters.
Several trends emerge from the parameter correlations (Fig. \ref{figtwo}). The rise and decay times are strongly correlated;  the photon index during GRBBLs is almost always harder than the long-term average, consistent with a harder-when-brighter behavior commonly reported in individual blazars. Most intriguingly, there is an anti-correlation between GRBBL luminosity and timescales: the brightest GRBBLs are the fastest, while low-luminosity GRBBLs tend to evolve more slowly.
This timescale–luminosity relation, although admittedly it shows a huge scatter, deserves to be investigated further, because it might open the window for cosmological studies with GRBBLs.\\

This work represents the first automated, population-level study of $\gamma$-ray blazar flares, or GRBBLs for short. By combining a robust identification pipeline with a simple exponential rise/decay model and unsupervised classification, I have characterized 679 GRBBLs (in the fiducial test, and up to 1572 in the test with looser quality cuts), revealing a largely homogeneous population with a small fraction of chromatic outliers. The work shows that population studies of GRBBLs are achievable with current tools, and provide new information, complementary to deep studies of individual events. Future efforts will focus on extending this analysis to finer time binnings, incorporating adaptive light-curve segmentation, and adding multi-wavelength information to the parameter space to search for additional sub-classes and outliers.\\

\bibliographystyle{iaulike} 
\bibliography{Sample} 

\end{document}